\documentclass[usenatbib,twocolumn]{mn2e}
\usepackage{graphicx}
\usepackage{amsfonts}
\usepackage{epsfig,rotating,natbib,pstricks}
\usepackage{color}
\usepackage{amssymb,latexsym}
\usepackage{amsmath,wasysym}

\pubyear{2010}

\def\beq{\begin{equation}}
\def\eeq{\end{equation}}   
\def\baq{\begin{eqnarray}}
\def\eaq{\end{eqnarray}}

\def\p3m{P$^3$M}
\def\ap3m{AP$^3$M}

\def\h1{H\/I}
\def\omegah1{\Omega_{\h1}}
\def\ph1{P_{_{\h1}}}
\def\ph1k{P_{_{\h1}}(k)}
\def\dh1k{\Delta^2_{_{\h1}}(k)}
\def\lfir{L_{\mathrm{FIR}}}
\def\lsun{L_{\odot}}
\def\lbol{L_{\mathrm{bol}}}
\def\msun{M_{\odot}}
\def\lco{L^{\prime}_{\mathrm{CO}(1-0)}}

\def\lbox{L_{\mathrm{box}}}
\def\npart{N_{\mathrm{part}}}
\def\hmpc{h^{-1}\mathrm{Mpc}}
\def\hkpc{h^{-1}\mathrm{kpc}}
\def\mdm{m_{_{\mathrm{DM}}}}
\def\mgas{m_{_{\mathrm{gas}}}}
\def\mbh{M_{_{\mathrm{BH}}}}
\def\mcold{M_{_{\mathrm{cold}}}}
\def\mmol{M_{_{\mathrm{mol}}}}
\def\mstar{M_{*}}
\def\mhalo{M_{\mathrm{halo}}}
\def\mblack{{\emph{MassiveBlack }}}
\def\lsim{\apprle}
\def\gsim{\apprge}
\def\tmax{T_{max}}

\newcommand{\be}{\begin{equation}}
\newcommand{\e}{\end{equation}}

\title[Quasar Hosts at $z\sim6$]{The Formation of Galaxies Hosting $z\sim6$ Quasars }

\author[Khandai et al.]{\parbox{18cm}{
    Nishikanta Khandai$^{1}$, 
    Yu Feng$^{1}$,
    Colin DeGraf$^{1}$,
    Tiziana Di Matteo$^{1}$, 
    Rupert A.C. Croft$^{1}$}\vspace{0.3cm}\\%,
  $^1$ {McWilliams Center for Cosmology, 
    Carnegie Mellon University, 5000 Forbes Avenue, Pittsburgh, PA 15213, USA}}
\pubyear{2011}
\label{firstpage}

\def\LaTeX{L\kern-.36em\raise.3ex\hbox{a}\kern-.15em
    T\kern-.1667em\lower.7ex\hbox{E}\kern-.125emX}

\pagerange{\pageref{firstpage}--\pageref{lastpage}} 

\begin{document}

\maketitle
%%%%%%%%%%%%%%%%%%%%%%%%%%%%%%%%%%%%%%%%%%%%%%%%%%%%%%%%%%%%%%%%%%%%%%%

\begin{abstract}
  We investigate the formation and properties of galaxies hosting $z \sim 6$
  quasars, in the gigaparsec scale cosmological hydrodynamical simulation:
  \mblack, which includes a self-consistent model for star formation, black
  hole accretion and associated feedback. We show that the \mblack reproduces
  current estimates of the galaxy stellar mass function $z=5, 6$. We find
  that quasar hosts in the simulation are compact gas rich systems with high
  star formations rates of SFR $\sim 100-10^3 \msun{\mathrm{yr}}^{-1}$
  consistent with observed properties of Sloan quasar hosts in the redshift
  range $5.5 \lsim z \lsim 6.5$.  We show that the star-forming gas in these
  galaxies predominantly originates from high density cold streams which are
  able to penetrate the halo and grow the galaxy at the center. \mblack
  predicts a deviation from the local $\mbh-\sigma$ and $\mbh-\mstar$ relation
  implying that black holes are relatively more massive for a given stellar
  host at these redshifts.
  
\end{abstract}

%%%%%%%%%%%%%%%%%%%%%%%%%%%%%%%%%%%%%%%%%%%%%%%%%%%%%%%%%%%%%%%%%%%%%%%

 
%%%%%%%%%%%%%%%%%%%%%%%%%%%%%%%%%%%%%%%%%%%%%%%%%%%%%%%%%%%%%%%%%%%%
\section{Introduction}

Supermassive black holes (SMBH) are now ubiquitously found in the nuclei of
local galaxies.  Tight correlations have been observed between the central
black hole and its host galaxy
\citep{1998AJ....115.2285M,2000ApJ...539L...9F,2000ApJ...539L..13G,2001ApJ...563L..11G,2002ApJ...574..740T,
  2004ApJ...604L..89H,2003ApJ...589L..21M}, implying that the growth of the
quasar, powered by the SMBH, is intimately linked to the formation of its host
galaxy.  At $z \sim 6$ and above the most direct constraint on the evolution
of SMBHs comes from observations of luminous quasars in the Sloan Digital Sky
Survey \citep{2006AJ....132..117F,2009AJ....138..305J}. 
Recently a detection has been confirmed for a quasar at $z=7$ \citep{2011Natur.474..616M}. 
These quasars are mainly optically selected and 
represent the bright end of the quasar population at this epoch. 
They are rare with number densities of $n \sim $ a few Gpc$^{-3}$ and
extremely luminous with inferred masses of $\mbh \sim 10^9 \msun$, suggesting
that they are harboured in rare halos of mass $\mhalo \sim 10^{13} \msun$ at
these redshifts.  Observations in other bands, far infrared (FIR)-radio,
suggest that the spectral energy distributions (SED) of these quasars are
similar to those of their low redshift counterparts
\citep{2008ApJ...687..848W}. This means that $10^9 \msun$ black holes are
\emph{fully developed} and already in place even when the Universe was
relatively young ($\lsim 10^9$ years at $z \gsim 6$).

Reprocessed thermal dust continuum emission in the FIR provides a clean method
for deriving the total star formation rates in galaxies
\citep{2002ApJ...576..159D}.  Bright quasars also add to the FIR luminosity,
$\lfir$, and one needs to correct for it to estimate the star formation rate
(SFR) of the host galaxy.  The excess $\lfir$ (corrected) for a sample of $z
\sim 6$ quasar hosts suggests a massive starburst origin (SFR $\sim 10^{2.7} -
10^{3.4} \msun$ yr$^{-1}$) for them
\citep{2010ApJ...714..699W,2011AJ....142..101W}. These authors also detected
large reservoirs of molecular gas, with mass $\mmol \sim 10^{10}\msun$, in
these galaxies through the emission of redshifted carbon monoxide (CO) line,
further corroborating the result that the observed large star formation
activity in these galaxies are sustained through an abundant supply of
molecular gas, the fuel for star formation. The cold gas in these galaxies is
localised within a radius of a few kpc
\citep{2004ApJ...615L..17W,2009Natur.457..699W,2010ApJ...714..699W}.  The
relation between CO luminosity, $\lco$, and $\lfir$ for these galaxies are
similar to those of typical star-forming systems at lower redshift.

However the relation between the black hole mass, $\mbh$ and the bulge
velocity dispersion of its host, $\sigma$, is seen to be above the local
relation\citep{2010ApJ...714..699W}, indicating that the $\mbh-\sigma$
relation is evolving. This is true even when some of the assumptions, like
degeneracy of $\sigma$ with inclination angle are considered. Even though
there is still significant debate whether there may be observational biases
influencing these results \citep{2007ApJ...670..249L}.

Recent observations with the refurbished Hubble Space Telescope (HST) have
yielded a considerably larger sample of more \emph{typical} galaxies at $ 5
\lsim z \lsim 8$. The newly installed Wide Field Camera (WFC3/IR) and the
Advanced Camera for Surveys (ACS) on the HST have been used to detect much
fainter galaxies out to $z \sim 8$. These provide strong constraints on the
UV luminosity function of galaxies
\citep{2011ApJ...737...90B,2011arXiv1105.2038B} at these redshifts. In
addition WFC3/IR+ACS in combination with the Spitzer telescope has been used
to detect the highest redshift ($z \sim 10$) galaxy till date
\citep{2011Natur.469..504B,2011arXiv1105.2297O}. The UV luminosity function
at these redshifts has been used to estimate the galaxy stellar mass
function (GSMF) from $z \sim 5-7$ \citep{2011ApJ...735L..34G}.

These observations of high redshift galaxies impose strong constraints on
theoretical models of galaxy formation. Hydrodynamic simulations (with models
for star formation) have been carried out for studying the global properties
of high redshift galaxies
\citep{2011MNRAS.414..847S,2011arXiv1104.2345J}. They are able to reproduce
the observed UV luminosity of these galaxies with some assumptions for
extinction due to dust.  However the shape of the GSMF
\citep{2011arXiv1104.2345J} is found to be inconsistent with those inferred
from observations \citep{2011ApJ...735L..34G}.  These simulations do not model
the growth of black holes and their feedback on the surrounding environment
and are typically \emph{small} in volume being incapable to host rare
high-sigma peak objects.

The requirement for large volumes and significant resolution to follow galaxy
formation (gas inflows into relatively small scales) has made numerical
studies of the growth of the first quasars extremely challenging. A number of
approaches have resorted to 'constrained' simulations. In a pioneering study,
\citep{2007ApJ...665..187L} high-resolution merger simulations with subgrid
models for star formation and growth of black holes have been used to study
the formation of $z \sim 6$ quasars. This work extracted merger trees from large,
coarse dark matter only simulations and identified the most massive halo
candidate at $z=6$. To simulate at high resolution the merger trees were
populated with isolated galaxies which undergo the corresponding mergers and
are then endowed with gas and models for star formation and black hole
growth.

This approach qualitatively reproduces the properties of the quasar SDSS
J1148+5251 and its host at $z=6.42$. Fully cosmological re-simulations of
selected halos from the Millenium run were also carried out in a more recent
study by \cite{2009MNRAS.400..100S}. This approach also finds that the most massive
halos are consistent with being the first quasar hosts.

In order to study directly how and where the first quasars assemble, without a
pre-imposed choice of halo, large cosmological volumes ($\sim \mathrm{Gpc}^3$)
for capturing rare high-sigma peaks are required and  sufficiently
high-resolution in order to resolve kpc scales. High resolution is also
necessary when including subgrid models for star-formation, black hole
accretion and related feedback processes. Keeping these constraints in mind we
have run the largest (currently feasible) hydrodynamic simulation of its kind,
\mblack, which includes gravity, hydrodynamics and subgrid models for
star-formation, black hole growth and associated feedback processes.  It was
run in a cosmological volume of $\lbox = 533 h^{-1}\mathrm{Mpc}$ with $2\times
3200^3 \approx 65$ billion particles (dark matter + gas) and a uniform
gravitational softening of $\epsilon = 5 h^{-1}\mathrm{kpc}$, with the code
{\small P-GADGET} which has been extensively modified from the public code
{\small GADGET2} \citep{2005MNRAS.364.1105S} to run optimally on a large
number of multicore processors.  The simulations were carried out on $10^5$
cores of Kraken at NICS\footnote{http://www.nics.tennessee.edu}.  \mblack has
the same mass and force resolution (as well as similar subgrid models for star formation,
black hole growth and associated feedback processes) as that of the resimulated halo of
\cite{2007ApJ...665..187L}, however it does not rely on an imposed merger scenario to
produce luminous quasars at $z \sim 6$ and tracks the assembly of galaxies in
a more self-consistent manner. 

The high redshift observations of galaxies, quasars and their hosts can be
used to test standard models of galaxy formation in previously unexplored
regimes.  Indeed \mblack has been instrumental in reproducing a number of
observational properties of high-z quasars,e.g.  the formation and abundances
of Sloan Digital Sky Survey (SDSS) 
type black holes of mass $\mbh \sim 10^9 \msun$ at $z=6$
\citep{2011arXiv1107.1253D} and also statistical properties, such as their
luminosity function and high-redshift clustering \citep{2011arXiv1107.1254D}.
In this work we study the formation and properties of galaxies hosting $z \sim
6$ quasars and as a sanity check we see how well observed global properties of
galaxies compare with those in \mblack.  Our paper is organised as follows. We
start with by describing the \mblack simulation in section~\ref{sec_methods}.
In section~\ref{sec_global_props} we compare the GSMF in \mblack with
observations and earlier simulations.  We next identify
a sample of potential quasar host galaxies at $z \sim 6$ and look at their
formation and growth in sections~\ref{subsec_growth} and
\ref{subsec_coldflow}.  In section~\ref{subsec_obs} we compare properties of
these galaxies in \mblack with recent observations.  We present our
conclusions in section~\ref{sec_conclusion}.

\begin{table}
      \begin{center}
        \begin{tabular}{c|c|c|c|c}
          \hline
          $L_{\mathrm{box}}$ & $N_{\mathrm{part}}$ & $\mdm$ & $\mgas$ & $\epsilon$ \\
          $\left(h^{-1}\mathrm{Mpc}\right)$& & $\left(h^{-1}\msun\right)$ & $\left(h^{-1}\msun\right)$ & $\left(h^{-1}\mathrm{kpc}\right)$ \\
          \hline
          533.333 & $2\times 3200^3$ & 2.78$\times 10^8$ & 5.65$\times 10^7$ & 5  \\
         \hline
        \end{tabular}
      \end{center}
\caption{Basic simulation parameters for the MB simulation. The columns
  list the size of the simulation box, $L_{\mathrm{box}}$, the
  number of dark matter particles used in the simulation, $N_{\mathrm{part}}$, 
  the mass of a single dark matter particle, $\mdm$, 
  the initial mass of a gas particle, $\mgas$, 
  and the gravitational softening length,
  $\epsilon$.  All length scales are in comoving units.}
\label{table_simparam}
\end{table}

\section{Methods: The Massive Black Simulation}
\label{sec_methods}
In this section we describe a large hydrodynamic simulation, \mblack, 
which we have run to study the high redshift Universe.
We have used {\small P-GADGET}, a significantly upgraded version of
{\small GADGET3} (see \citep{2005MNRAS.364.1105S} for an earlier version)
 which we are developing for use at
upcoming Petascale supercomputer facilities.
\mblack is a cosmological simulation of a $\Lambda$CDM cosmology. 

It is worth pointing
out that the number density of $z\sim6$ quasars is extremely small,
$n \sim$ a few $(\mathrm{Gpc})^{-3}$, and that they
are hosted by rare massive halos.
Therefore in order to simulate and resolve these objects one needs a large
cosmological volume as well as high mass and force
resolution.  \mblack was run with $\npart
= 2\times3200^3 = 65.5$ billion particles in a comoving volume of side $\lbox
= 533\hmpc$ and a comoving gravitational softening length of
$\epsilon=5\hkpc$, (see table~\ref{table_simparam} for more details).

The initial conditions were generated with the Eisenstein and Hu power 
(\citation)
spectrum at $z=159$ and the simulation was evolved to $z=4.75$.  The
cosmological parameters used were: the amplitude of mass fluctuations,
$\sigma_8=0.8$, spectral index, $n_s = 0.96$, cosmological constant parameter
$\Omega_{\Lambda}= 0.74$, mass density parameter $\Omega_m = 0.26$ and baryon density 
parameter $\Omega_b = 0.044$.

Along with gravity and smoothed particle hydrodynamics (SPH) {\small P-GADGET}
incorporates a multi-phase ISM model with star formation
\citep{2003MNRAS.339..289S} and black hole accretion and feedback
\citep{2005Natur.433..604D,2005MNRAS.361..776S}.

\subsection{Subgrid model for black hole accretion and feedback}
\label{sec_global_props}
\begin{figure*}
\begin{tabular}{c}
  \includegraphics[width=7.0truein]{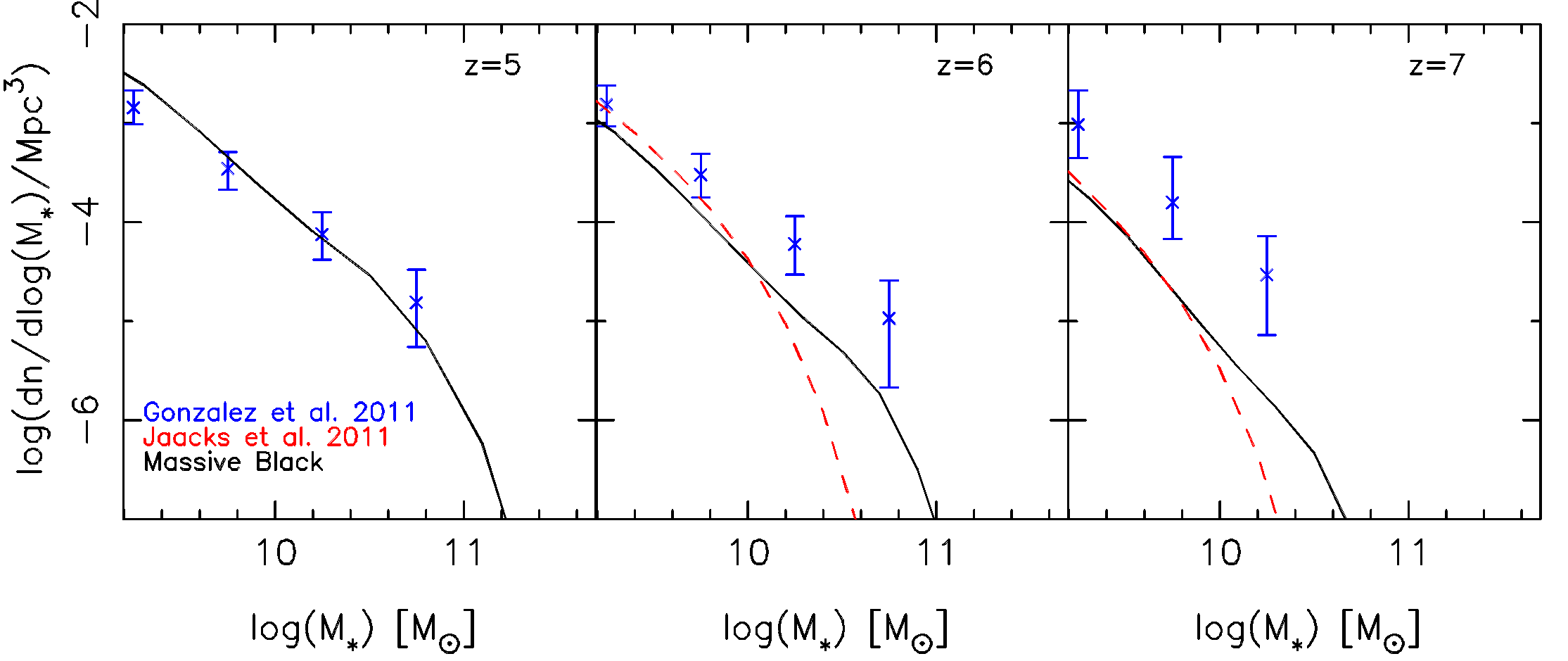} \\
\end{tabular}
\caption{The evolution of the GSMF from $z=5$ to $z=7$ (left to right) in the MB simulation (solid line), 
data points are from observations reported by \protect\cite{2011ApJ...735L..34G}. 
The dashed curve is from hydrodynamic simulations of high redshift galaxies 
by \protect\cite{2011arXiv1104.2345J}.}
\label{gsmf}
\end{figure*}

Black holes are modelled as collisionless sink particles within newly
collapsing halos in our simulation, which are identified by a
friends-of-friends (FOF) \citep{1985ApJ...292..371D} halofinder called on the
fly at regular time intervals.  A seed black hole of mass $M_{\mathrm{seed}} =
5\times10^{5}h^{-1}\msun$ is inserted into a halo with mass $\mhalo \geq
5\times10^{10}h^{-1}\msun$ if it does not already contain a black hole.  The
seeding recipe is chosen to match the expected formation of SMBHs by gas
directly collapsing to black holes with $\mbh \sim M_{\mathrm{seed}}$
\citep{2003ApJ...596...34B,2006MNRAS.370..289B} or by massive primordial stars
collapsing into $\sim 10^2 \msun$ mass black holes
\citep{2004ARA&A..42...79B,2006ApJ...652....6Y} at $z \sim 30$.  Once seeded,
black holes grow by accreting surrounding gas or by merging with other black
holes.  Gas is accreted with an accretion rate $\dot{M}_{\mathrm{BH}} =
\frac{4\pi G^2 \mbh^2 \rho}{\left(c_s^2 + v^2\right)^{3/2}}$
\citep{1939PCPS...34..405H,1944MNRAS.104..273B,1952MNRAS.112..195B}, where
$\rho$ is the local gas density, $c_s$ is the local sound speed and $v$ is the
relative velocity of the black hole and the surrounding gas. We limit the
accretion rate to mildly super-Eddington consistent with
\cite{2006MNRAS.370..289B,2006ApJ...650..669V} to prevent artificially high
values. The black hole radiates with a bolometric luminosity which is
proportional to the accretion rate, $\lbol = \eta\dot{M}_{\mathrm{BH}}c^2$
\citep{1973A&A....24..337S}, where $\eta$ is the radiative efficiency and its
standard value of 0.1 is kept throughout, and $c$ is the speed of light. Some
of the liberated energy is expected to couple thermally
 to the surrounding gas.  In the
simulation 5\% of the radiated energy does this.
This energy is deposited isotropically on gas particles
that are within the black hole
kernel (32 nearest neighbours) and acts as a form of feedback energy
\citep{2005Natur.433..604D}. The value of 5\% is the only parameter in the
model and was set using galaxy merger simulations\citep{2005Natur.433..604D}
to match the normalisation in the observed $\mbh - \sigma$ relation.  Black
holes also grow by merging with other black holes once they come within the
spatial resolution with a relative velocity below the local gas sound speed.

This model for the growth of black holes has been developed by
\cite{2005Natur.433..604D,2005MNRAS.361..776S}.  It has been implemented and
studied extensively in cosmological simulations
\citep{2007MNRAS.380..877S,2007ApJ...665..187L,2008MNRAS.387.1163C,2008ApJ...676...33D,
  2009MNRAS.400...43C,2009MNRAS.398...53B,2009MNRAS.400..100S,2010MNRAS.402.1927D,
  2011MNRAS.413.1383D,2011MNRAS.tmp.1136D,2011arXiv1104.3550C} successfully
reproducing basic properties of black hole growth, the observed $\mbh -
\sigma$ relation and the black hole mass function \citep{2008ApJ...676...33D},
the quasar luminosity function \citep{2010MNRAS.402.1927D} as well as the
clustering of quasars \citep{2011MNRAS.413.1383D}.

We identify galaxies with the group-finder code {\small SUBFIND} \citep{2001MNRAS.328..726S}.
{\small SUBFIND} identifies bound clumps within a FOF halo
and properties of galaxies such as position, mass (dark matter, gas, stars and black holes), 
star formation rate (SFR) among others are stored. 

We use a relational database management system developed by \cite{2011lopez}
specifically for this simulation to track the history of black hole
properties  (e.g. mass, accretion rate, position, local gas density,
sound speed, velocity, and black hole velocity relative to local
gas) which are saved for each black hole at every timestep.
For a complete summary of the database format and its efficiency, 
the reader is referred to \cite{2011lopez}.

\section{Results}
\subsection{The Galaxy Stellar Mass Function}
We start by looking at the global properties of galaxies at high redshift in
the \mblack simulation. We use the halo catalogues generated by SUBFIND and
pick all halos with more than 40 star particles to construct a galaxy stellar
mass function (GSMF) from $z=7$ to $z=5$.  This translates to a minimum
stellar mass of galaxies of $M_* = 10^{9.2}\msun$.  We compare our GSMF with
recent observed mass functions \citep{2011ApJ...735L..34G} which were derived
from the luminosity function at these redshifts
\citep{2011ApJ...737...90B}. We also compare our GSMF with those from recent
hydrodynamic simulations at these redshifts by \cite{2011arXiv1104.2345J}.  We
note that \cite{2011arXiv1104.2345J} used multiple runs with varying
resolution to achieve a larger dynamic range in resolved galaxies, and were
run with the same SPH code 
and cosmological parameters as the \mblack run in this
study.  Feedback due to AGNs was however not included by
\cite{2011arXiv1104.2345J}.  Two examples of the Jaack et al.  runs
are the N600L100 ($\lbox =
100h^{-1}\mathrm{Mpc}$, $\npart=2\times600^3$) of \cite{2011arXiv1104.2345J}
which has exactly the same mass resolution as our \mblack simulation and their
N400L10 run ($\lbox = 10h^{-1}\mathrm{Mpc}$, $\npart=2\times400^3$) 
which can probe galaxies masses
down to $M_* = 10^{6.8}\msun$.

Our results (solid line) are shown in figure~\ref{gsmf} for $z=5,6$ and $7$
(left to right).  The data points are from observations at these redshifts
\citep{2011ApJ...735L..34G}.  The dashed lines are from
\cite{2011arXiv1104.2345J} and were only computed down to $z=6$.  At $z=5$ our
results are in good agreement with observations in the mass range that we are
able to resolve.

At $z=6$ we slightly underpredict the GSMF for $M_* > 10^{9.5}\msun$, but
a main difference with \cite{2011arXiv1104.2345J} is
that their GSMF curve drops off completely for $M_* >
10^{10.5}\msun$. This may be due to the smaller volume ($\sim 150$ times
 smaller than MB) that they have used.
However they are consistent with our estimates
for $M_* < 10^{10}\msun$.

At $z=7$ both the theoretical GSMFs underpredict that of the observations in
the entire mass range.  The source of this discrepancy is at present
unclear. This may be due to the relation between mass and light used by
\citep{2011ApJ...735L..34G} which was calibrated at $z \sim 4$ and is
consistent with the sample at $z \sim 5$. At $z \sim 6$ and $z \sim 7$ this
correlation is less obvious even though it is consistent, in zero point, with
the even smaller sample at these higher redshifts. Our GSMF are evolving but
the observations are consistent (within error-bars) with no evolution in the
GSMF from $z=5$ to $z=7$.  We will explore these issues as well as the
luminosity function of high redshift galaxies, which is the direct observable
in a future work.

\section{Properties of Host Galaxies of $z \sim 6$ Quasars}
We now focus on individual properties of host galaxies of  $z\sim6$ quasars
in this section. We look at how these galaxies and the 
black holes at their center were assembled and compare our results with their observed 
properties in the latter part of this section. 

\subsection{Star formation and black hole growth}
\label{subsec_growth}
\begin{figure*}
\begin{tabular}{c}
  \includegraphics[width=7.0truein]{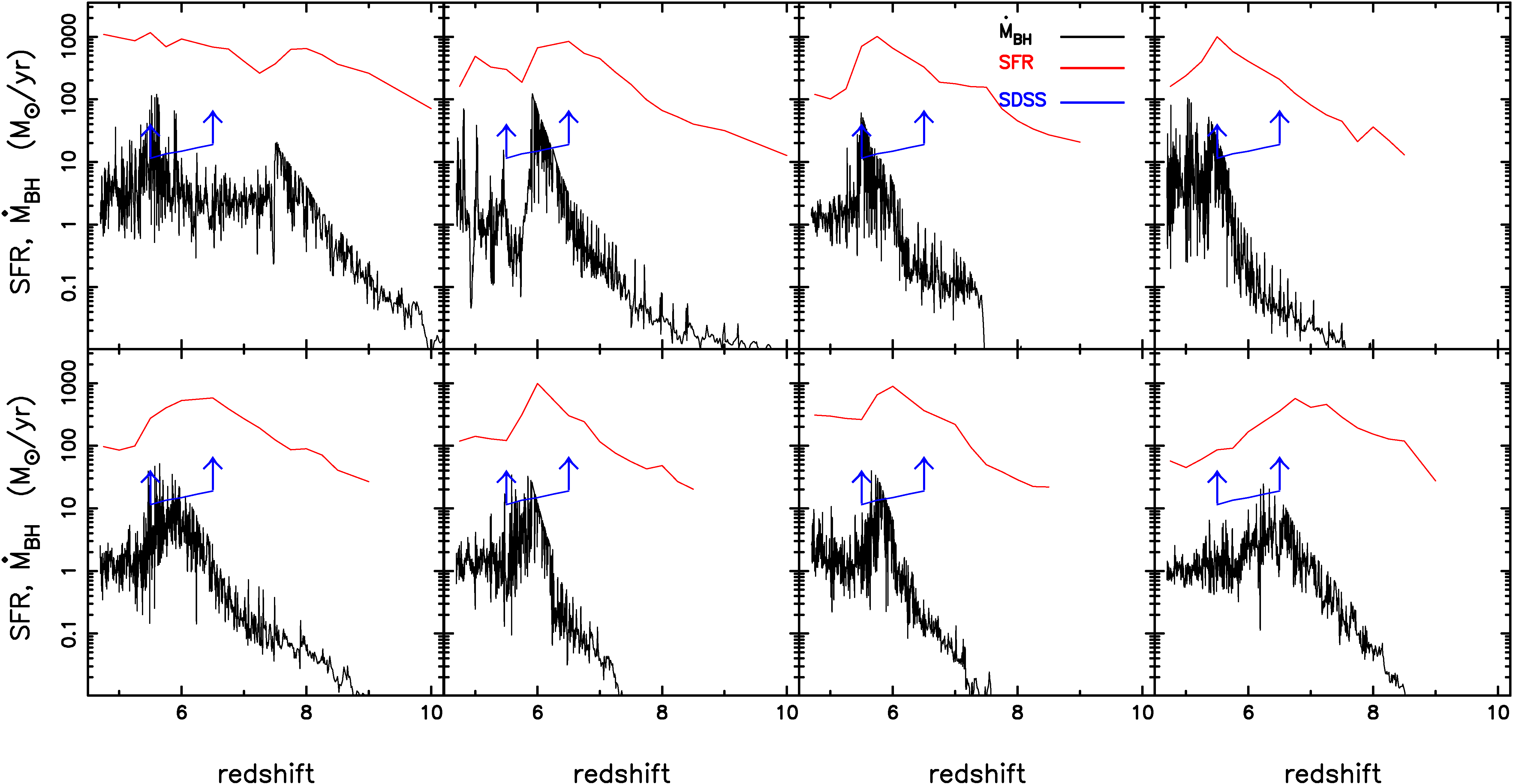} \\
\end{tabular}
\caption{SFR(red) and black hole accretion rate(black) in $\msun$/yr. 
The SDSS flux limit of $m_i < 20.2$ for $z > 3$
\protect\citep{2009ApJ...697.1656S} for the quasar sample is shown by a blue line.}
\label{fig_sfrhalo}
\end{figure*}

Our simulation allows us to follow the growth of supermassive black holes and
their host galaxies up to $z \sim 5$.  In Figure~\ref{fig_sfrhalo}, we plot
the accretion rate of the black hole (black) and the SFR (red) of its host
with redshift for a sample of eight luminous quasars selected such that their
luminosity is consistent with the magnitude limit for quasars in the Sloan
survey. This is the blue line, which is bounded 
by an $i$-band magnitude limit of $m_i <
20.2$ for $z > 3$ \citep{2009ApJ...697.1656S} and converted to a bolometric
luminosity (and then an accretion rate) using the SED of
\cite{2007ApJ...654..731H}. This SDSS flux cut roughly corresponds to
BH accretion rates of about 10 solar masses per year). 

\begin{figure*}
\begin{tabular}{ccccccc}
  \includegraphics[width=1.04truein]{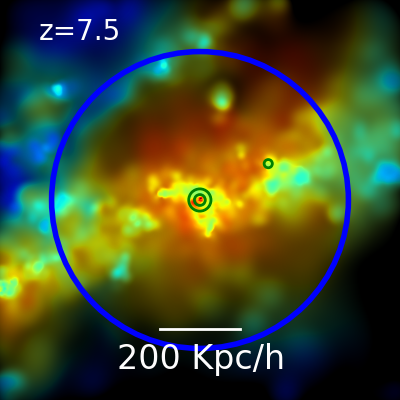} 
  \includegraphics[width=1.04truein]{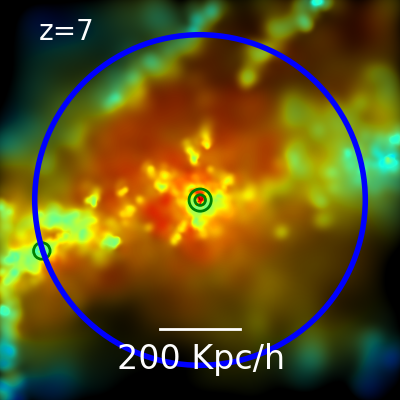}
  \includegraphics[width=1.04truein]{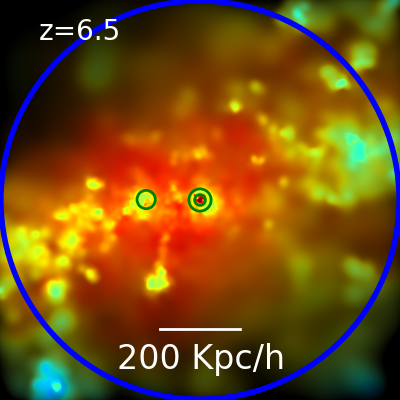}
  \includegraphics[width=1.04truein]{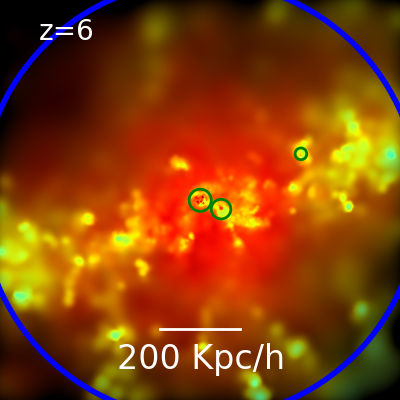}
  \includegraphics[width=1.04truein]{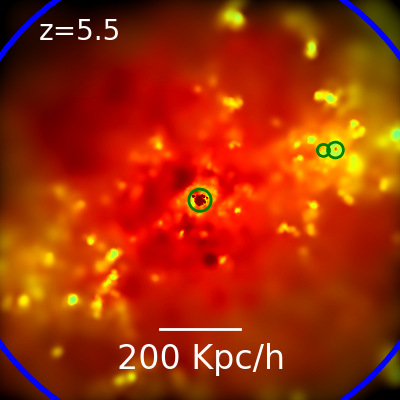}
  \includegraphics[width=1.04truein]{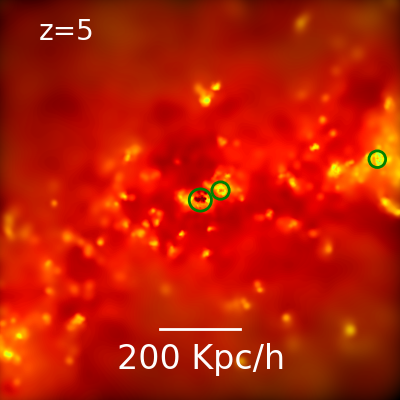}
  \includegraphics[width=1.04truein]{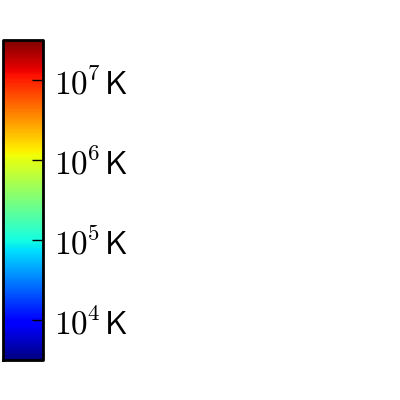}\\
  \includegraphics[width=1.04truein]{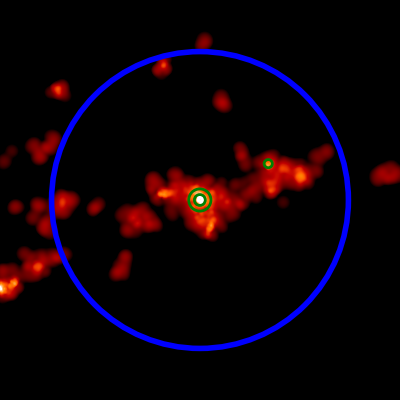} 
  \includegraphics[width=1.04truein]{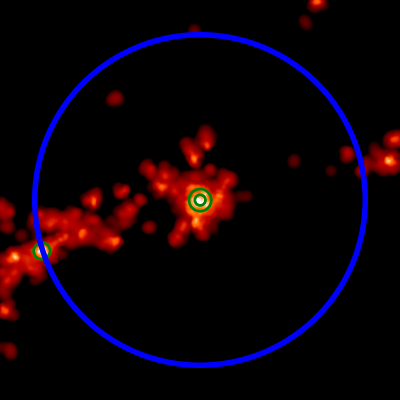}
  \includegraphics[width=1.04truein]{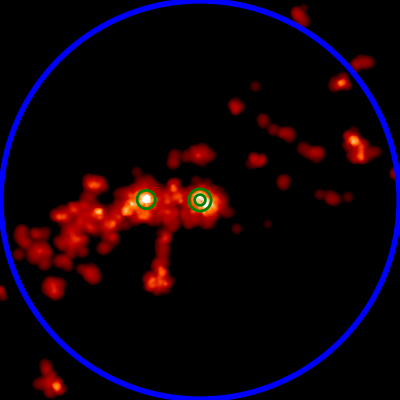}
  \includegraphics[width=1.04truein]{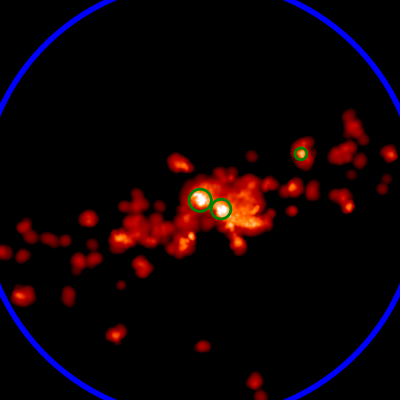}
  \includegraphics[width=1.04truein]{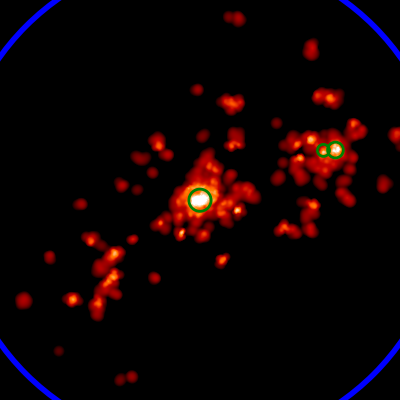}
  \includegraphics[width=1.04truein]{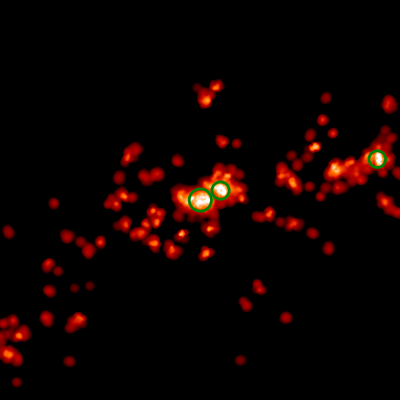}
  \includegraphics[width=1.04truein]{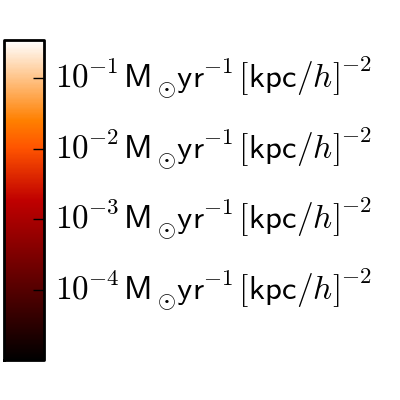}\\
  \includegraphics[width=1.04truein]{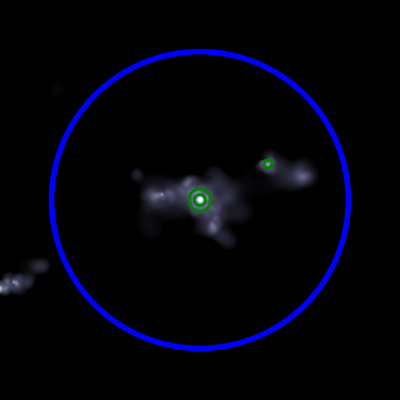} 
  \includegraphics[width=1.04truein]{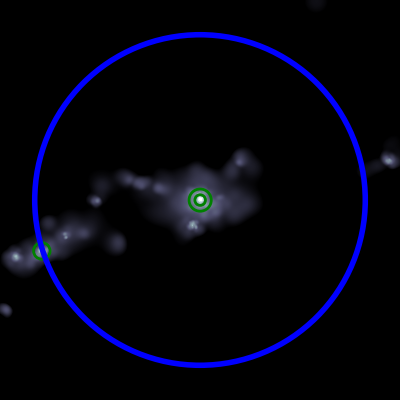}
  \includegraphics[width=1.04truein]{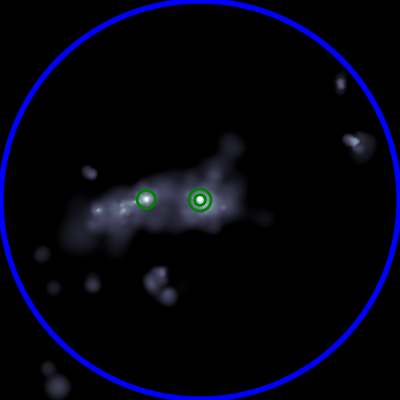}
  \includegraphics[width=1.04truein]{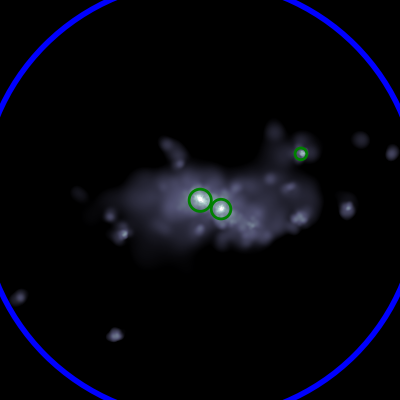}
  \includegraphics[width=1.04truein]{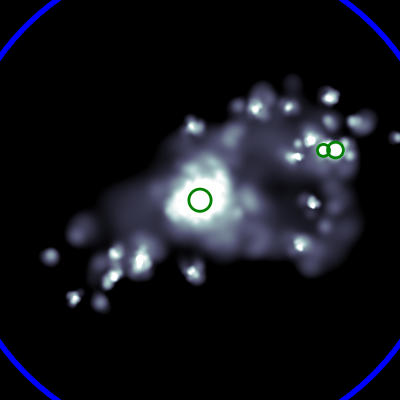}
  \includegraphics[width=1.04truein]{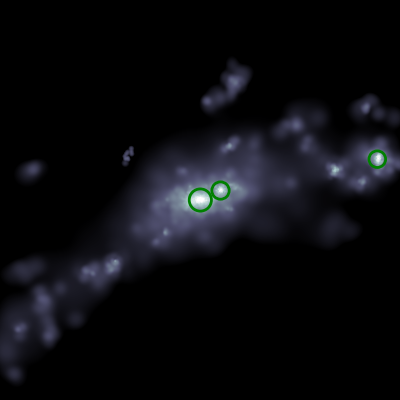}
  \includegraphics[width=1.04truein]{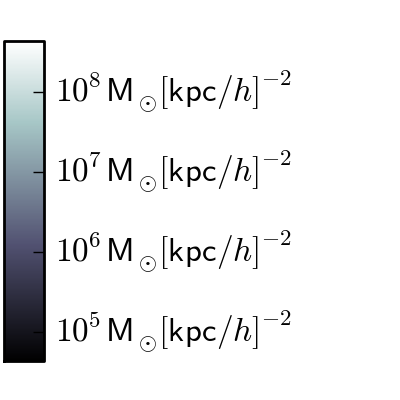}\\
\end{tabular}
\caption{An example of the growth of a typical $z \sim 6$ quasar host galaxy. This host corresponds to 
the first  object in the top panel of figure~\ref{fig_sfrhalo}. 
The top row visualizes the gas distribution color coded by temperature across six redshifts.
The projected density ranges from $\sim 10^{5}$ to $\sim 10^{8.5} \msun$ (kpc/h)$^{-2}$.
In the middle row we visualize the gas distribution but now color coded
by the SFR.
The bottom row shows the distribution of stars.
The large blue circle is the virial radius of the halo.
The smaller circles are black holes with the radius proportional to their mass.}
\label{fig_tempsfr}
\end{figure*}

In figure~\ref{fig_sfrhalo} we find that the 
star formation rate in the galaxy is
strongly correlated with the growth of the central black hole. The 
central black
holes grow rapidly through a period of sustained Eddington accretion (
typically between $8 <z < 6$) and are continuously fed by streams of high
density gas \citep{2011arXiv1107.1253D}.  The mass of the host halos in this sample grow 
from $\mhalo \sim 10^{11.6}$  at $z = 8$ to $\mhalo \sim 10^{12.4}$  at $z = 6$ .   
Prior to its peak accretion phase
the black hole grows more rapidly that the stellar mass in its host. Star
formation is regulated by feedback from the black hole, and typically star
formation is suppressed just prior to the peak accretion phase of the black
hole. This is a typical feature of this model and has been seen in many
previous works \citep{2005Natur.433..604D,2007ApJ...665..187L,2009MNRAS.400..100S}.
%{\bf some   hopkins papers}. 
Once the black hole accretion becomes feedback dominated it deposits
enough energy in its vicinity and shuts off further growth by expelling gas in
its surrounding star-forming region.  At its peak, the SFR of the host galaxy
is extremely high, $\mathcal{O}(10^3) \msun/\mathrm{yr}$ , and is consistent
with observations \citep{2010ApJ...714..699W,2011AJ....142..101W} for the SFR
of quasar hosts at these redshifts. We make a direct
 comparison with observations in
a later section.  Most of the SMBHs fall within the Sloan detection limit when
they grow through Eddington limited accretion and attain $\mbh = 10^9 \msun$
during this phase. The one exception is the first object which peaks early on
at $z=7.5$ then undergoes a merger (as will become evident in
figure~\ref{fig_tempsfr}) at $z=5.5$ to become a $\mbh = 10^9
\msun$ black hole.

\begin{figure*}
\begin{tabular}{ccccccc}
  \includegraphics[width=1.04truein]{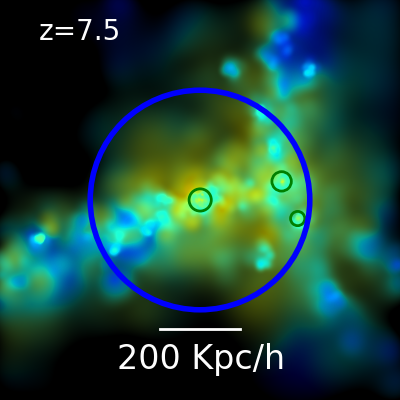} 
  \includegraphics[width=1.04truein]{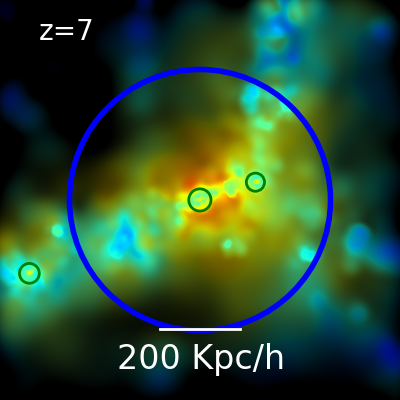}
  \includegraphics[width=1.04truein]{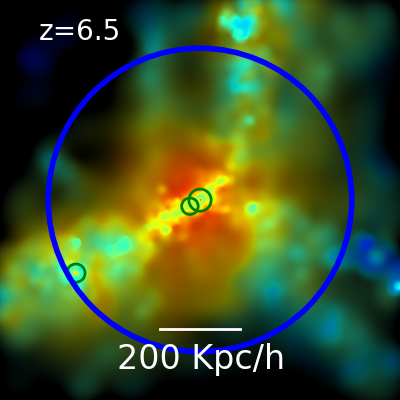}
  \includegraphics[width=1.04truein]{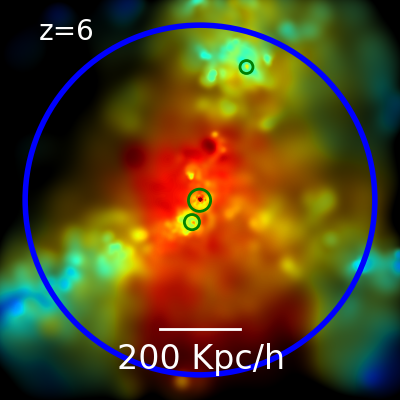}
  \includegraphics[width=1.04truein]{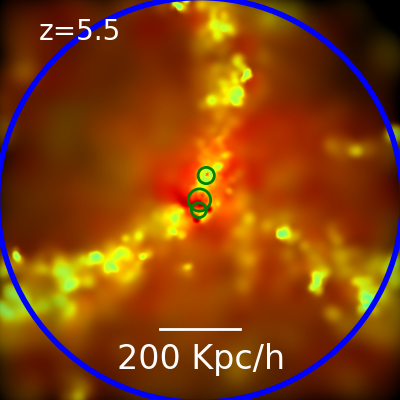}
  \includegraphics[width=1.04truein]{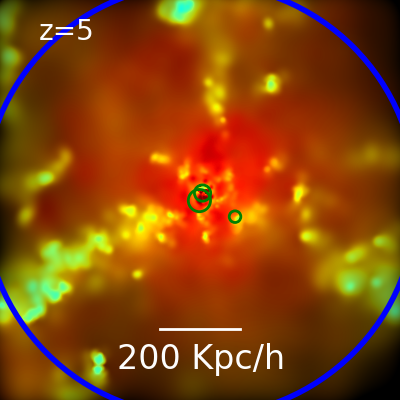}
  \includegraphics[width=1.04truein]{temp-colorbar.png}\\
  \includegraphics[width=1.04truein]{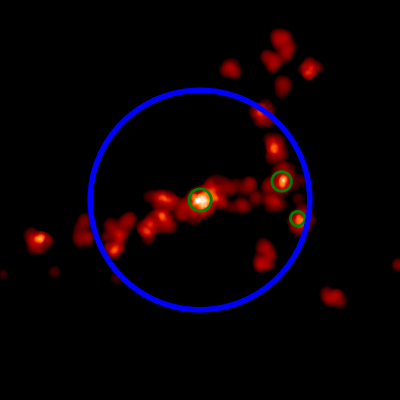} 
  \includegraphics[width=1.04truein]{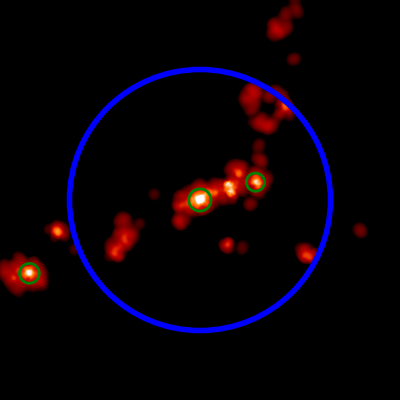}
  \includegraphics[width=1.04truein]{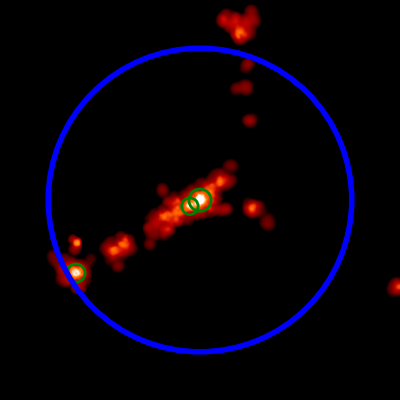}
  \includegraphics[width=1.04truein]{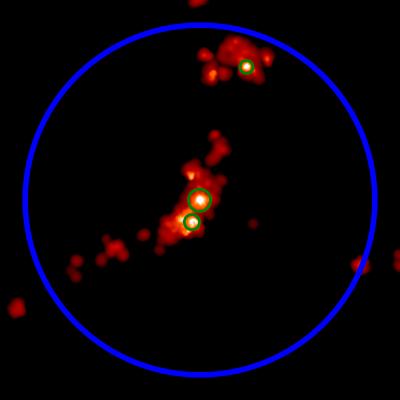}
  \includegraphics[width=1.04truein]{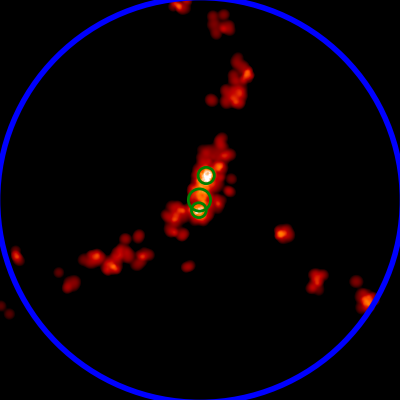}
  \includegraphics[width=1.04truein]{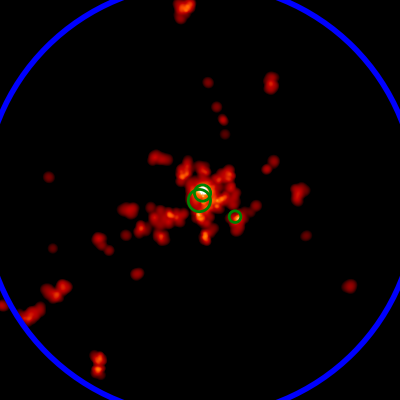}
  \includegraphics[width=1.04truein]{sfr-colorbar.png}\\
  \includegraphics[width=1.04truein]{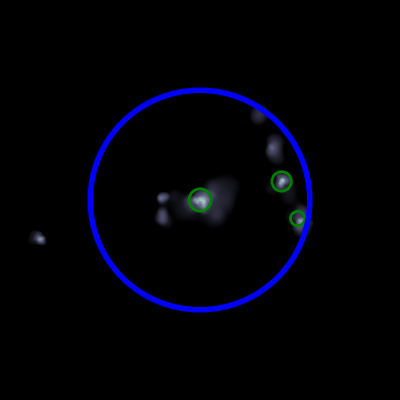} 
  \includegraphics[width=1.04truein]{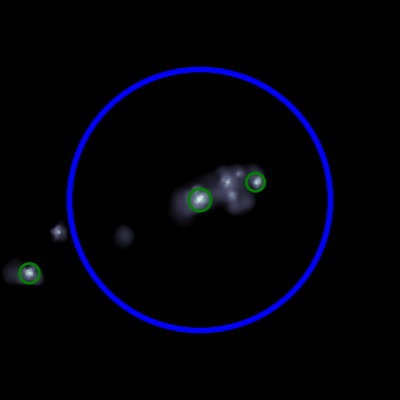}
  \includegraphics[width=1.04truein]{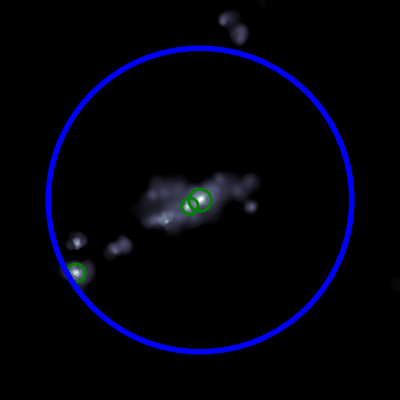}
  \includegraphics[width=1.04truein]{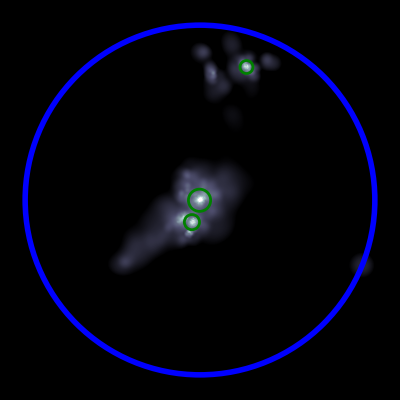}
  \includegraphics[width=1.04truein]{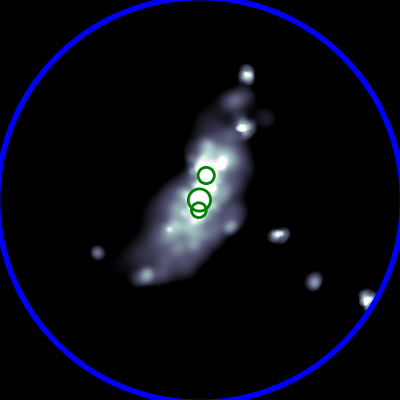}
  \includegraphics[width=1.04truein]{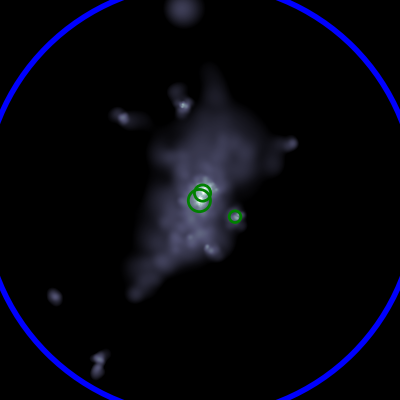}
  \includegraphics[width=1.04truein]{star-colorbar.png}\\
\end{tabular}
\caption{Same as in figure~\ref{fig_tempsfr} but for the second object, top row, in figure~\ref{fig_sfrhalo}.}
\label{fig_tempsfr2}
\end{figure*}

\begin{figure*}
\begin{tabular}{cccccc}
  \includegraphics[width=1.04truein]{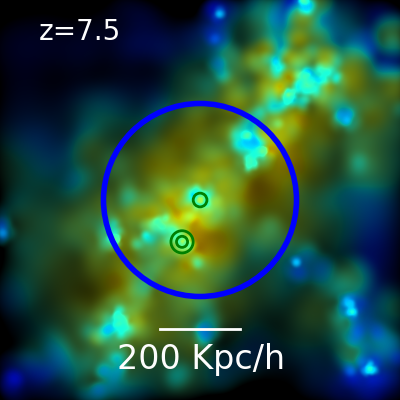} 
  \includegraphics[width=1.04truein]{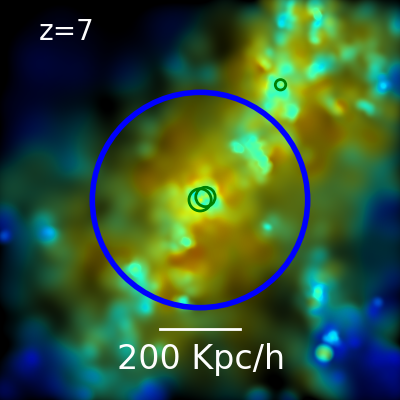}
  \includegraphics[width=1.04truein]{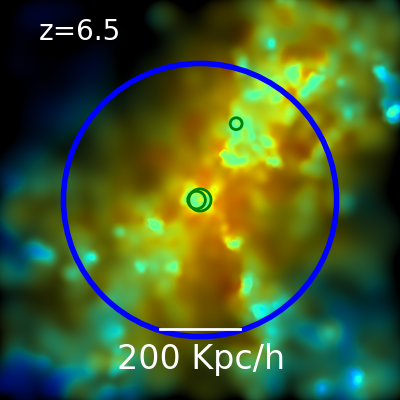}
  \includegraphics[width=1.04truein]{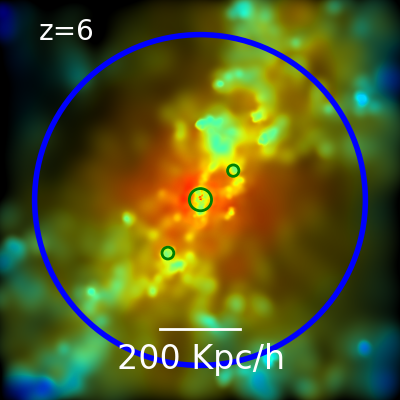}
  \includegraphics[width=1.04truein]{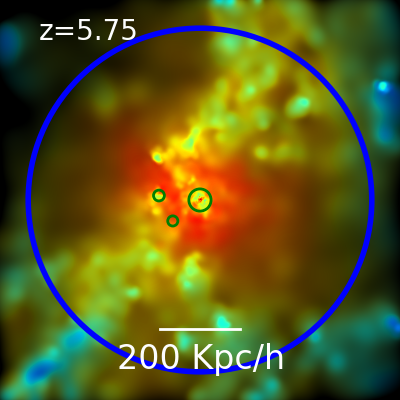}
  \includegraphics[width=1.04truein]{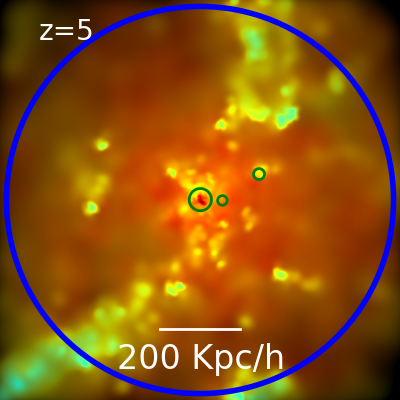}
  \includegraphics[width=1.04truein]{temp-colorbar.png}\\
  \includegraphics[width=1.04truein]{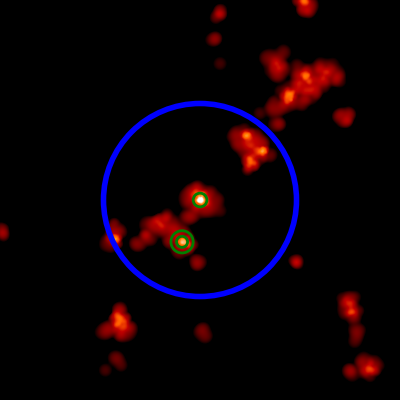} 
  \includegraphics[width=1.04truein]{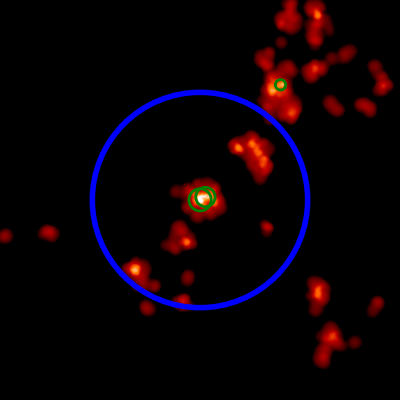}
  \includegraphics[width=1.04truein]{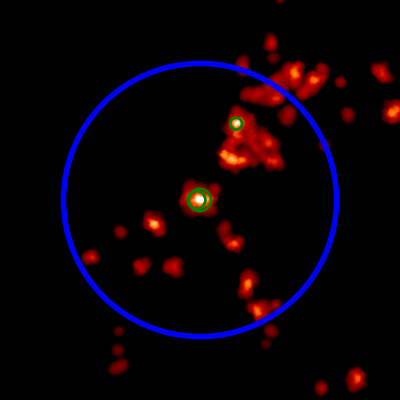}
  \includegraphics[width=1.04truein]{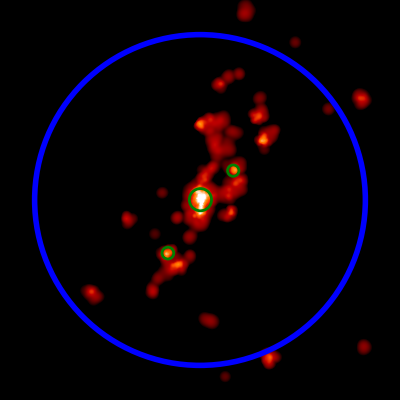}
  \includegraphics[width=1.04truein]{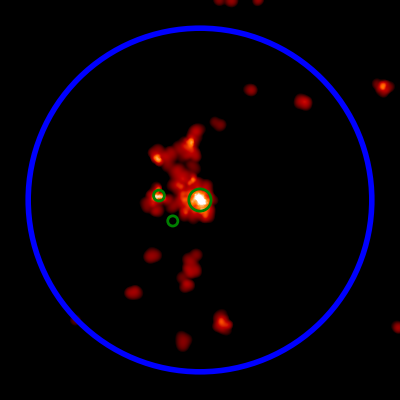}
  \includegraphics[width=1.04truein]{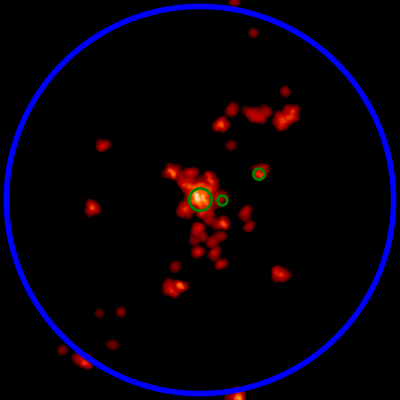}
  \includegraphics[width=1.04truein]{sfr-colorbar.png}\\
  \includegraphics[width=1.04truein]{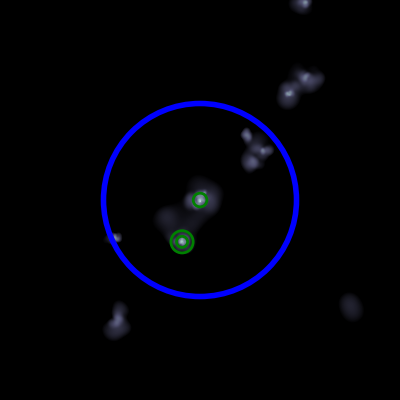} 
  \includegraphics[width=1.04truein]{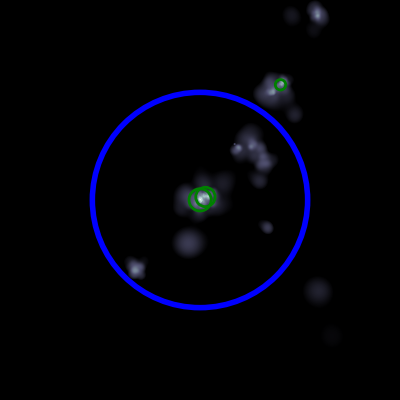}
  \includegraphics[width=1.04truein]{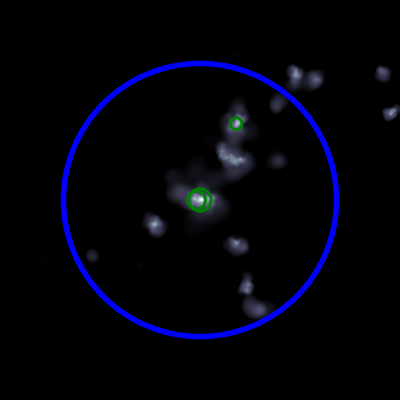}
  \includegraphics[width=1.04truein]{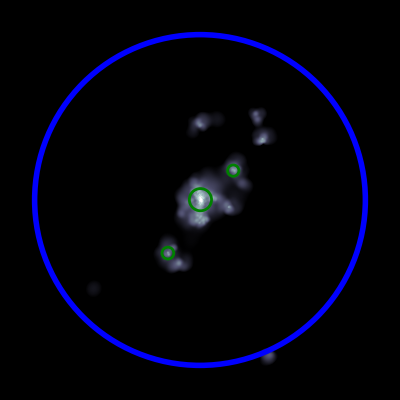}
  \includegraphics[width=1.04truein]{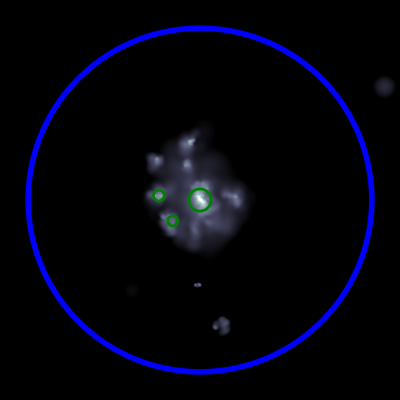}
  \includegraphics[width=1.04truein]{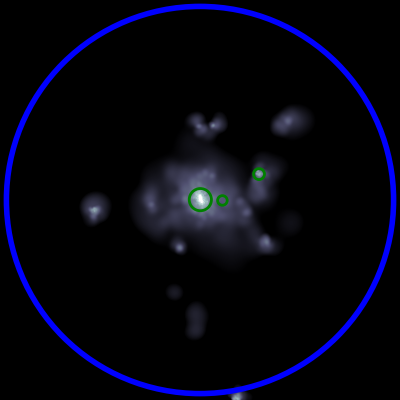}
  \includegraphics[width=1.04truein]{star-colorbar.png}\\
\end{tabular}
\caption{Same as in figure~\ref{fig_tempsfr} but for the third object, top row, in figure~\ref{fig_sfrhalo}.}
\label{fig_tempsfr3}
\end{figure*}

We look at the evolution of the environment around three typical host galaxies
(the first three objects in the top row of figure~\ref{fig_sfrhalo}) from $z=7.5$
to $z=5.0$ in figures~\ref{fig_tempsfr}-\ref{fig_tempsfr3}.  In the top row of
these figures we plot the gas distribution, color coded by temperature.  The
middle row is the gas distribution color coded by the SFR and the bottom panel
is the distribution of stars. The large blue circle denotes the virial radius
of the halo and the smaller circles are black holes with the radius
proportional to the mass of the black hole. 

The SMBH in these examples have different growth histories, e.g. the first
SMBH becomes feedback dominated at $z=7.5$ then grows again due to a merger at
$z=5.5$ seen in figure~\ref{fig_tempsfr}.  Apart from this SMBH the others
grow to a mass of $10^9 \msun$ by Eddington limited accretion by $z \sim 6$.
As seen in figures~\ref{fig_tempsfr2}-\ref{fig_tempsfr3} these halos are
continuously fed by cold streams down to $z \sim 6$, prior to the feedback
dominated phase.  At this redshift feedback from the black hole starts to
inject energy but is not able to disrupt the stream.  Star formation is still
sustained at $\mathrm{SFR} \sim 10^3 \msun/\mathrm{yr}$ further down to $z
\sim 5.5$.  By $z=5.5$ feedback from the black hole starts to destroy the
inner structure of the cold streams, inhibiting further growth; the SFR drops
down by an order of magnitude of its peak value. There are still pockets of
cold star-forming gas around the black hole, however the depleted gas due to
star formation is no longer replenished by the cold streams.  By $z=5$ the
inner part of the smooth cold stream is mostly destroyed, the gas in the
central region is heated to $T\sim 10^7$K and pushed out and
dense subhalos which
have been self-shielded from feedback contribute to the growth of the host
galaxy and the central black hole, though at a much reduced rate.

In the middle row of figures~\ref{fig_tempsfr}-\ref{fig_tempsfr3} we again
look at the distribution of gas around the same object but now color coded by
the SFR. The redshift of each panel is the same as in the top row.  We find
that the central region of the galaxy is forming most of the stars.  Star
formation also occurs in dense clumps of gas located on cold filamentary streams 
though at a much reduced rate. The central region has a sustained
period of star formation down to $z=5.5$ for most objects.  During this period
the black hole has grown at the Eddington rate through smooth accretion of
cold gas. From this inspection it appears that most star formation also occurs
in the most centrally-located gas that also feeds the black hole which, as demonstrated
previously \citep{2011arXiv1107.1253D} is cold-flow fed.

At $z=5$ the star formation has dropped drastically in the central galaxy due
to feedback from the central black hole.  It appears that feedback eventually
destroys the cold streams inside the halos; star formation still persists
though at a reduced level in dense clumps surrounding the black hole.  In the
bottom row of figures~\ref{fig_tempsfr}-\ref{fig_tempsfr3} we look at the
distribution of stars.  As expected the locations of stars are preferentially
found near star-forming gas particles (middle row).  We are unable to
accurately predict the morphologies of the galaxies due to the limited
resolution of the simulation.

\subsection{Growth of Host Galaxies Through Cold Streams}
\label{subsec_coldflow}
\begin{figure*}
\begin{tabular}{c} 
  \includegraphics[width=7.0truein]{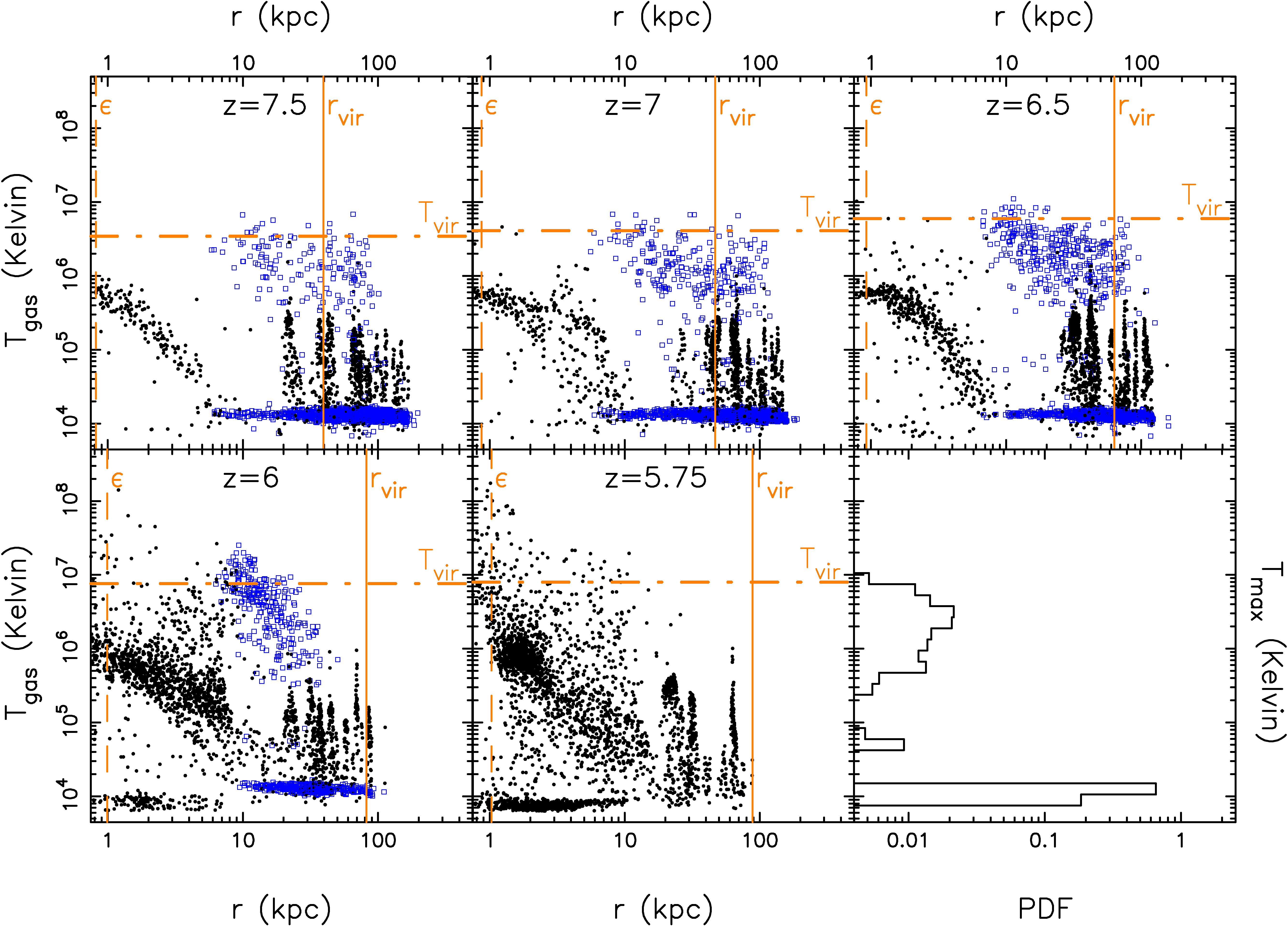}
\end{tabular}
\caption{Temperature histories of the star-forming gas of the central halo at $z = 5.75$ as a function of 
separation from the central SMBH. All length scales in this figure are in physical units.   
This particular halo is the same 
as in figure~\ref{fig_tempsfr} and has a peak SFR at $z\sim5.75$. 
Star forming gas particles in the central halo are tracked backwards 
with respect to the reference redshift of $z=5.75$ when this halo was at its peak
star-forming activity.
The horizontal dot-dashed line denotes the virial temperature (Kelvin) of the halo, the solid horizontal
line is the virial radius of the halo and the dashed horizontal line is the gravitational softening length.
The filled black circles denote gas particles with non-zero SFR and the open blue squares 
are gas particles with zero SFR. 
The bottom right panel indicates the distribution of $\tmax$ for star-forming gas particles at $z=5.75$}
\label{fig_tdr}
\end{figure*}

Here we investigate the origin of the star forming gas in the quasar
host galaxies. We will in particular test whether most of this
star forming gas is indeed entering the halo via cold streams rather than cooling from
a shock heated phase. We wish to point out that we do not characterise 
the redshift of accretion of these star-forming gas particles, but rather look at the origin 
of all the star-forming gas during the host galaxy's peak star-forming activity.

%This is indeed what is expected for halos in this range of mass
%from the work of \cite{2005MNRAS.363....2K,2006MNRAS.368....2D,2009MNRAS.395..160K},
%where one expects the hot mode accretion to dominate for halos with mass 
%$\mhalo > 10^{11.4}\msun$. 

As an example, we look at the object in figure~\ref{fig_tempsfr3}.  We examine
how the star-forming gas in the halo at its peak star-forming activity,
i.e. at $z=5.75$, was accreted.  We trace the temperature histories of these
star-forming gas particles back to $z=7.5$ and plot them as a function of
(physical) separation from the SMBH, in figure~\ref{fig_tdr}.  To make
comparison with observations easier, in the discussion that follows all length
scales are quoted in physical units. The temperatures represent the effective
temperature for star-forming gas particles which are in the two-phase medium
\citep{2003MNRAS.339..289S} and the temperature for non star-forming gas which
are in the single phase medium.

In figure~\ref{fig_tdr} the filled black circles denote star-forming gas
particles and the open blue squares denotes the gas particles which have zero
SFR. The horizontal dot-dashed line is the virial temperature, the vertical
solid line is the virial radius and the vertical dashed line is the
gravitational softening length.  Prior to $z=5.75$, one can identify four
distinct regimes in the $T-r$ plots of the gas particles which end up forming
stars in the halo: (1) a cold stream of non-star forming gas with temperatures
of $\sim 10^4$K beyond $r \sim 10$kpc, (2) clumps of star-forming gas outside
$r \sim 10$kpc with temperatures $10^4 \lsim T \lsim 10^6$K, (3) hot non-star
forming gas outside $r \sim 10$kpc with temperatures $10^{5.5} \lsim T \lsim
10^7$K and finally (4) dense star-forming gas within $r \sim 10$kpc with
temperatures $10^4 \lsim T \lsim 10^6$K.  Star formation only occurs in dense environments,
i.e. at the centre of the halo and clumps located in filaments, this is also
seen in figures~\ref{fig_tempsfr}-\ref{fig_tempsfr3}.  A further investigation
reveals that the cold non star-forming gas (with $T \sim 10^4$K) is also
located in filaments whereas the hot non star-forming gas is diffuse and
spread out across the halo.

As can be seen in figure~\ref{fig_tdr} most of the gas that is star-forming
at $z=5.75$ does not come from the diffuse hot medium.  
The major mode of gas accretion for the host galaxy is
from gas in filamentary streams.  The streams have two components, dense star
forming clumps and cold non star-forming gas that penetrates deep into the
halo well within the 10 kpc region of the central black hole. At this point gas
is dense enough to form stars in the central galaxy and also fuel the growth
of the black hole. The mass of gas in star forming clumps in the stream is
small compared to the cold non-star forming gas in the stream. 
Feedback from supernovae and the
black hole heats up the gas at the centre, so that
the temperature systematically
increases to $T \sim 10^6$K with decreasing separation.  This mode of
accretion continues down to $z \sim 5.75$.

Eventually by $z=5$ the black hole has injected enough energy into the
surrounding medium to destroy the cold stream supplying gas to the central
galaxy. The black hole's growth has become self-regulated. The black
hole has also regulated the growth of its host.  We find little
star-formation within 3 kpc of the  black hole, but a residual amount of
dense gas is still clumped in the region with  
$3 \lsim r \lsim$ 10 kpc, so that star-formation still
persists in this part of the central galaxy.

A sustained period of cold accretion is  largely
responsible for the high SFR for most objects. 
The exception is the first object
in figure~\ref{fig_sfrhalo} where a merger at $z=5.5$ is responsible for 
increasing the SFR.  
As seen in figures~\ref{fig_tempsfr}-\ref{fig_tempsfr3} star formation occurs only in
dense regions, i.e. mostly in the central halo and at a reduced level in dense
clumps within filaments.  At its peak, $95\%$ of star formation occurs within
the 3 kpc region of the central galaxy.  
The numbers quoted for this example are representative of those for
the full sample, fluctuating by only a few percent for the
other galaxies.

We track the entire temperature history of the star-forming gas particles of
the host galaxy at its peak star-forming phase, i.e. at $z=5.75$.  
We define $\tmax$ as the
maximum temperature that a gas particles attains prior to its first
star-forming phase, i.e. prior to it first being in a two-phase medium
\citep{2005MNRAS.363....2K,2009MNRAS.395..160K}.  A distribution of $\tmax$
will then indicate whether these gas particles were accreted through the hot
or cold mode \citep{2005MNRAS.363....2K,2009MNRAS.395..160K}.  This is shown
in the bottom right panel of figure~\ref{fig_tdr}. 
As can be seen the majority ($> 85\%$) of the
star-forming gas is accreted through the cold mode, the remaining gas ($<
15\%$) comes from gas which has been heated while accreting onto the halo and
then cools to form stars.  These \emph{hot} gas particles are the same as the diffuse hot
non star-forming gas seen in the first four panels of figure~\ref{fig_tdr}. 
The numbers presented here represent a lower bound for cold mode accretion for star-forming gas,
since a larger fraction of star-forming gas particles prior to $z=5.75$ should have 
accreted onto the halo through the cold mode. 
We find that if we considered all the gas, not distinguishing between 
star-forming and non star-forming gas, then $\sim 70\%$ of the gas at $z=5.75$ is accreted 
cold and $\sim 30\%$ is accreted through the hot mode. 

The analysis presented here indicates that the quasar host galaxies are
forming stars very efficiently.
The massive star formation is sustained
through a supply of gas mostly from cold streams, that are able to penetrate
the halo and reach the center without being heated to the virial temperature
of the halo consistent with earlier work at lower redshift \citep{2009Natur.457..451D}. 
A smaller fraction of gas $< 15\%$ does get heated to $\sim
T_{\mathrm{vir}}$ to finally cool and form stars. 
At its peak $\sim 95\%$ of the star-formation is occurring within a 3 kpc region of the central halo,
consistent with observations \citep{2004ApJ...615L..17W,2009Natur.457..699W,2010ApJ...714..699W}.

\begin{figure*}
\begin{tabular}{cc}
  \includegraphics[width=3.3truein]{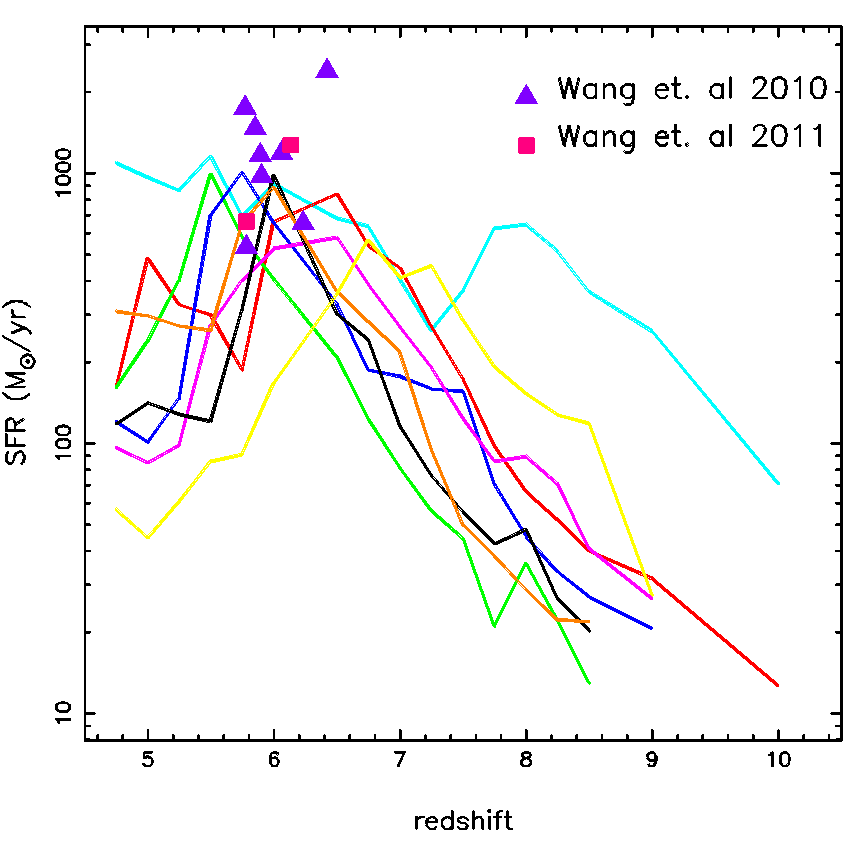} 
  \includegraphics[width=3.3truein]{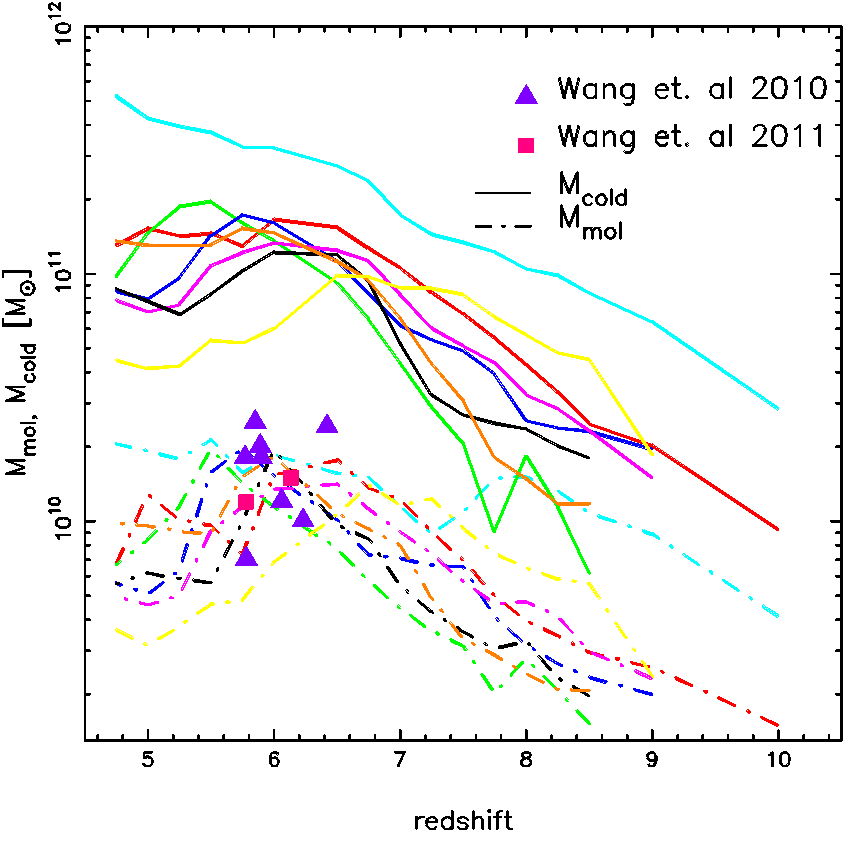} \\
\end{tabular}
\caption{\emph{Left}:Evolution of SFR for the hosts of the luminous quasars. 
\emph{Right}:Evolution of $\mcold$ (solid lines) and $\mmol$ (dot-dashed)
of the hosts of the most massive black holes. 
Data points in both panels 
are from observations \protect\citep{2010ApJ...714..699W,2011AJ....142..101W}. 
Data points on the right panel are estimates for $\mmol$.}
\label{fig_sfr}
\end{figure*}

\subsection{Comparison with Observations}
\label{subsec_obs}
In this section, we compare properties of quasar hosts in the \mblack
simulation with recent observations \citep{2010ApJ...714..699W}.  These
observations specifically constrain observables for the host galaxies such as
SFR, molecular gas $\mmol = M\left[\mathrm{H}_2+\mathrm{He} \right]$, and the
$\mbh - \sigma$ relation.  Additionally we look at the $\mbh - \mstar$
relation for the host galaxies in the \mblack simulation and compare it with
the local relation in observations.

\subsubsection{Star Formation Rates and Cold Gas}
The reprocessed emission in the far-infrared (FIR) from star formation-heated
dust is used to provide an estimate of the star formation
from observations of $z \sim
6$ quasar host galaxies \citep{2008ApJ...687..848W}.  The AGN contribution
is removed and the remaining
FIR luminosities, $\lfir$, are then  converted  to a SFR
\citep{2010ApJ...714..699W,2011AJ....142..101W}, using 
\beq 
\mathrm{SFR} \simeq
2.55 \times 10^{-10} \left(\msun \mathrm{yr}^{-1}\right)
\left(\frac{\lfir}{\lsun}\right). 
\label{eq_sfr2lfir}
\eeq

The sample of quasar host galaxies in the observations of
\cite{2010ApJ...714..699W,2011AJ....142..101W} fall in the redshift range
$5.78 \lsim z \lsim 6.43$.  We compile the SFR of the quasar hosts 
in our simulation (the ones
in Figure~\ref{fig_sfrhalo}) and compare it with the observations in the left
panel of Figure~\ref{fig_sfr}.  We find that most of these host galaxies have
a peak SFR $\sim 10^{3} \msun.\mathrm{yr}^{-1}$ in
the redshift range $5.5 \leq z \leq 6.5$,
comparable to the observed SFR of quasar host galaxies. However our model
seems to lack objects close to the observed SFR of $1500 \msun\mathrm{yr}^{-1}$ for some host galaxies.
It is of course possible that our star formation model is somewhat simplistic, particularly 
at these redshifts, since we do not model star formation from molecular gas
(e.g., \cite{2011ApJ...729...36K}). 
Further investigation of the star formation modelling
 is beyond the scope of this paper so instead we
 prefer to compare directly
to associated measurements of the cold and molecular gas,
 which is what the observations can constrain. 

In the right panel of figure~\ref{fig_sfr} we plot the cold gas mass, $\mcold$, of the host.
We find that the the evolution of $\mcold$ broadly follows the trend seen in the SFR. Observations 
on the other hand probe the molecular gas mass, $\mmol$ of these hosts which are not directly modeled
in our simulations.

The size of the cold molecular gas reservoir which fuels
star formation in these
observed quasar host galaxies has been estimated through redshifted CO
emission, \citep{2003A&A...409L..47B,2003Natur.424..406W,2007ApJ...666L...9C,2010ApJ...714..699W}.
These studies indicate molecular gas masses of $\gsim 10^{10}\msun$ in these
objects. Since our simulations do not model molecular gas instead estimate
the amount of molecular gas by using the SFR as a proxy for $\mmol$, 
i.e. by converting the SFR to $\lfir$, eq.~\ref{eq_sfr2lfir} and
using the relation between the CO (1-0) line luminosity, $\lco$, and $\lfir$
for local star-forming systems such as local starburst spiral galaxies, Ultra
Luminous Infra Red Galaxies (ULIRGs) and high-z submillimeter galaxies (SMGs)
\citep{2005ARA&A..43..677S}. 
\beq \log\left(\lfir\right) =
1.7\times\log\left(\lco\right) - 5
\label{eq_co2fir}
\eeq

\begin{figure*}
\begin{tabular}{cc}
  \includegraphics[width=3.3truein]{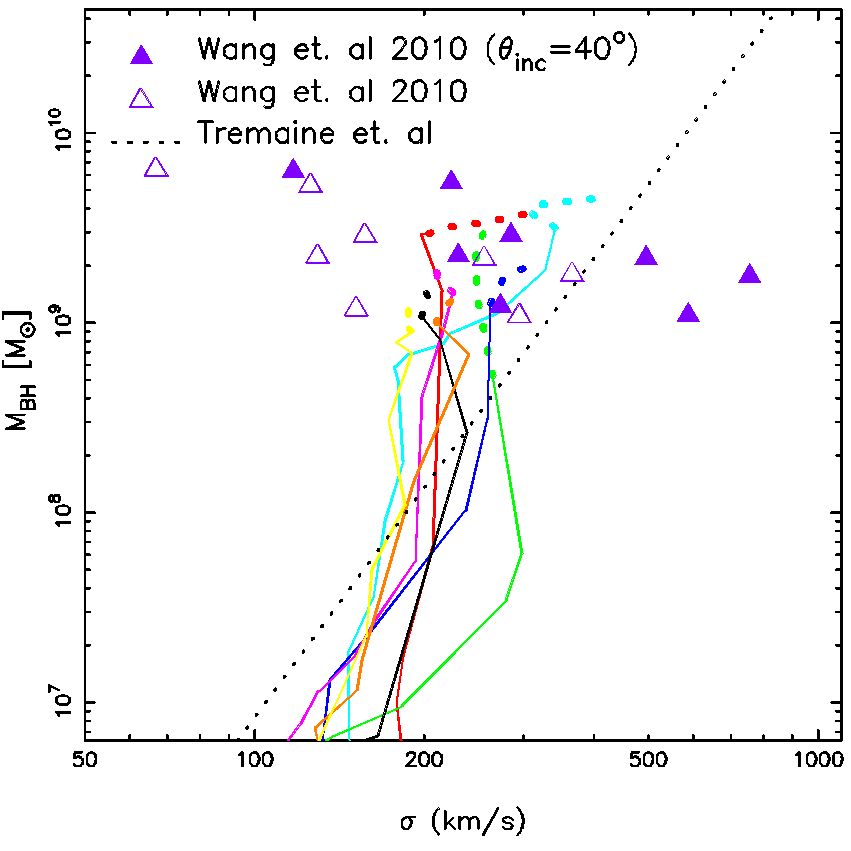} 
  \includegraphics[width=3.3truein]{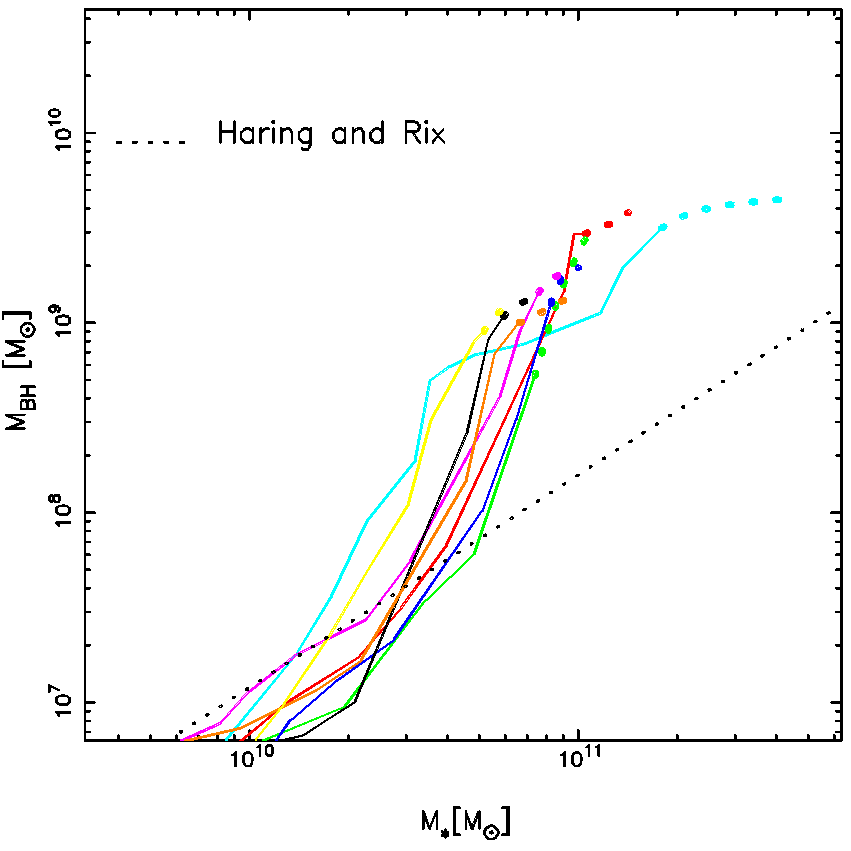} \\
\end{tabular}
\caption{Evolution of $M_{BH}-\sigma$ (left) for quasar-host systems  
in the \mblack simulation up to $z = 5.5$ (solid line) 
and further down to $z=4.75$ (dot-dashed line). 
The dotted line is the best fit local relation of 
\protect\cite{2002ApJ...574..740T}.
The filled triangles are data from observations 
$z \sim 6$ quasars and host galaxies \protect\citep{2010ApJ...714..699W} 
assuming an average inclination angle $\theta_{\mathrm{inc}} = 40^{\circ}$ and open triangles
are the same observations without any assumption for the inclination angles.
Evolution of the $M_{BH}-M_*$ relation (right) for the same objects as in the left panel.
Solid lines follow the evolution to $z=5.5$ and dot-dashed lines 
continue it down to $z=4.75$.
Comparison is made with the the best fit local relation of  \protect\cite{2004ApJ...604L..89H}.}
\label{fig_msigma}
\end{figure*}

$\lco$ can then be converted to a cold molecular gas mass $M_{\mathrm{cold}} =
\alpha \lco$ with $\alpha = 0.8 \msun
\left(\mathrm{K}\,\,\mathrm{km}\,\,\mathrm{s}^{-1}\,\,\mathrm{pc}^2\right)^{-1}$
\citep{2010ApJ...714..699W}.  Using these relations we look at the evolution
of  molecular gas for quasar host galaxies in the \mblack simulation and
compare them with observations \citep{2010ApJ...714..699W,2011AJ....142..101W}
in the right panel of figure~\ref{fig_sfr} shown as dot-dashed lines.  
Here again we find that we are able to reproduce the amount of
 molecular gas in these galaxies at
redshifts $5.5 \leq z \leq 6.5$. 
The estimates are in better agreement than
the observed SFR. \cite{2010ApJ...714..699W} also find that $z \sim 6$ quasar
hosts lie above the local $\lfir$-$\lco$ relation and attribute it to unknown
AGN contributions to $\lfir$ and different dust temperatures and CO line
ratios, even though they attempt to correct for AGNs.
 
The ratio between $\mmol$ and $\mcold$ is typically 10\%. This is much higher
than 1\% seen in observations of local galaxies \citep{2009MNRAS.394.1857O}.
This indicates that $z \sim 6$ quasar host galaxies are forming 
stars mores efficiently than local galaxies.

\subsubsection{The $\mbh-\sigma$ and the $\mbh-M_*$ relation}
We now look at the evolution of $\mbh-\sigma$ and the $\mbh-M_*$ relation 
between the central black hole and its host galaxy, where $\sigma$ is the velocity 
dispersion of the bulge, for which we use 
the velocity dispersion of stars within the half-mass radius as a proxy.
Given that the black hole grows faster than its host, as seen in figure~\ref{fig_sfrhalo}
we expect to see these relations to be steeper than the local relations, e.g.
\cite{2002ApJ...574..740T,2004ApJ...604L..89H}.

The evolution of the $\mbh-\sigma$ relation (solid line) is shown in the left
panel of figure~\ref{fig_msigma} for the \mblack sample.  Comparison is made
to the best fit local relation (dotted line) \citep{2002ApJ...574..740T} and
observations of $z \sim 6 $ quasar-host systems (triangles)
\citep{2010ApJ...714..699W}.  In this panel we show the evolution
up to $z=5.5$ (solid lines) in order to better compare with observations at these redshifts and further down to 
$z=4.75$ (dot-dashed lines).
Since the inclination angle is degenerate with the CO linewidths, an average 
inclination angle of $\theta_{\mathrm{inc}} = 40^{\circ}$ was assumed
(\cite{2010ApJ...714..699W}, filled triangles);  the open triangles relax
this 
constraint. 
As expected we see that the black holes grow more rapidly than their host
galaxies and eventually end up above the local relation of
\citep{2002ApJ...574..740T}.  By $z=5.5$ the $\mbh-\sigma$ relation compares
reasonably well with observations which assumes an inclination angle.  
However observations indicate that there
is an object with $\sigma \sim 120 $km/s and $\mbh \sim 10^{9.7}\msun$ even when the assumption 
of inclination angle is considered.  
We do not find such an object within our sample and this discrepancy 
may have to do with the uncertainties in the inclination angle.

Next we look at the evolution of the $\mbh-\mstar$ relation in the \mblack
simulation (solid lines) and compare with the best fit local relation of
\cite{2004ApJ...604L..89H} (dotted line) in the right panel of
figure~\ref{fig_msigma}. Again consistent with previous properties such as the
SFR-$\dot{M}_{\mathrm{BH}}$ relation and the $\mbh-\sigma$ relation we find
that the black hole is assembled more rapidly than the central host galaxy so
that the $\mbh-\mstar$ relation is well above the local relation of
\cite{2004ApJ...604L..89H}.  It would be interesting with 
future simulations to see what mechanisms
eventually drive them to the local relation.

The steep growth of $\mbh$ which outpaces the stellar component of the host galaxy
is common for all these systems and is again consistent with the fact that the accretion 
rate of the black hole as a function of redshift has a steeper slope than the SFR. 
The cyan and red lines in figure~\ref{fig_msigma} correspond to the first two objects 
of in the top row of figure~\ref{fig_sfrhalo}. A closer inspection shows that these objects 
undergo relatively dry mergers with subhalos in the feedback dominated phase. 
The subhalo has a larger stellar component compared to gas and therefore 
during this merger both $\sigma$ and $\mstar$ grow more rapidly than $\mbh$.
This can be seen 
as a relative flattening of the $\mbh-\sigma$ and $\mbh-\mstar$ relation in figure~\ref{fig_msigma}.
It is plausible that dry mergers are  responsible for eventually 
bringing the high redshift 
$\mbh-\sigma$ and $\mbh-\mstar$ relations onto the well constrained local relation 
\citep{2002ApJ...574..740T,2004ApJ...604L..89H}    

\section{Conclusions}
\label{sec_conclusion}
\mblack has been successful in reproducing a number of important properties of
quasars at high redshift within observational constraints: (i) the formation
and abundances of Sloan-type black holes of mass $\mbh \sim 10^9 \msun$ at
$z=6$ \citep{2011arXiv1107.1253D}, (ii) statistical properties such as the
luminosity function of quasars and its evolution as well as their
high-redshift clustering \citep{2011arXiv1107.1254D}. Importantly \mblack has
indicated that cold flows can sustain the growth of black holes at Eddington
rates so that they attain masses of $\sim 10^9 \msun$ in a short time span
\citep{2011arXiv1107.1253D}.

In this paper we have looked at the formation of galaxies hosting these $z \sim
6$ quasars with the \mblack simulation. We are able to reproduce a number of
observational properties of these galaxies.  We summarize our findings below:

\begin{itemize}
\item We find good agreement between the theoretical GSMF produced by \mblack
  and observations at $z=5$ and $z=6$. 
\item At $z \sim 6$, $\mbh \sim 10^9 \msun$ black holes are already in place.
  The growth of the black hole and its host galaxy are strongly correlated. 
  The black hole regulates the growth of its host galaxy. 
\item Black holes grow faster than the host galaxy and this is reflected 
  in the deviations seen in our study from the local $\mbh-\sigma$ relation 
  and  $\mbh-\mstar$ relation; our results are however consistent with observational 
  findings for these deviations at $z \sim 6$. 
\item The cold streams that feed the black hole 
  are also responsible for feeding the host galaxy
  to sustain the excessive starburst in them.  
  The accretion of gas, which eventually form stars in the host,
  through the \emph{hot mode} is subdominant ( $ \lsim 15\%$),
  the dominant fraction being accreted cold during their history.
\item Host galaxies grow in extreme environments. 
  We are able to reproduce the high SFR of $\sim 10^3 \msun\mathrm{yr}^{-1}$ 
  seen in observations in redshifts  $5.5 \lsim z \lsim 6.5$. Our simulation suggests 
  that the observed galaxies are at the peak of their star-formation activity 
  at these epochs.
\item Our derived estimates of molecular gas in these galaxies  
  are consistent with observations. Such large reservoirs of cold molecular gas, $\mcold \geq 10^{10}\msun$
  are responsible for fuelling star-formation in these galaxies.
\item Using the SFR as a proxy for $\mmol$ we find that the ratio 
  of molecular and cold gas is larger than seen for local galaxies. 
  This again indicates the host galaxies are forming stars more efficiently than 
  their local counterparts.
\item We find that most of the 
  star formation occurs within a compact 3-6 kpc (physical) region around the black hole at the peak of its 
  star-forming activity. These scales are again consistent with observations 
  \citep{2004ApJ...615L..17W,2009Natur.457..699W,2010ApJ...714..699W}.
\item Given that we are able to reproduce observations between $5.75 \lsim z \lsim 6.5$ 
  our simulations suggests that these quasars and their host galaxies are seen 
  at the peak of their assembly.
\end{itemize}

From the results presented here we expect that compared to the local Universe
the $\mbh-\sigma$ relation is very different in the high redshift Universe.
This has been indicated in earlier work \citep{2008ApJ...676...33D}, though
the sample was too small to put strong constraints on the relation at $z = 5$
and above.  Given the large volume and high resolution (hence large sample of
galaxies and black holes) \mblack is well suited for predicting the evolution
of the $\mbh-\sigma$ and $\mbh-\mstar$ relations at $z = 5$ and above.  It
would also be interesting to look at the mechanisms responsible for their
evolution.  We will address these issues in subsequent work.

\section*{Acknowledgments}
This work was supported by NSF award OCI-0749212. This research was supported by an
allocation of advanced computing resources provided by the National
Science Foundation.  The computations were performed on Kraken 
at the National Institute for Computational Sciences
(http://www.nics.tennessee.edu). NK would like to thank Volker Springel
for useful comments on the manuscript.

\label{lastpage}

\end{document}